\documentclass[prb,twocolumn,showpacs,amsmath,amssymb]{revtex4}

\usepackage{graphicx}
\usepackage{dcolumn}
\usepackage{bm}

\usepackage{color}

\begin{document}

\newcommand{\bn}{{\bf n}}
\newcommand{\bp}{{\bf p}}   
\newcommand{\br}{{\bf r}}
\newcommand{\bk}{{\bf k}}
\newcommand{\bv}{{\bf v}}
\newcommand{\brho}{{\bm{\rho}}}
\newcommand{\bj}{{\bf j}}
\newcommand{\wk}{\omega_{\bf k}}
\newcommand{\nk}{n_{\bf k}}
\newcommand{\eps}{\varepsilon}
\newcommand{\la}{\langle}
\newcommand{\ra}{\rangle}
\newcommand{\be}{\begin{eqnarray}}
\newcommand{\ee}{\end{eqnarray}}
\newcommand{\intl}{\int\limits_{-\infty}^{\infty}}
\newcommand{\dE}{\delta{\cal E}^{ext}}
\newcommand{\SE}{S_{\cal E}^{ext}}
\newcommand{\dsp}{\displaystyle}
\newcommand{\phit}{\varphi_{\tau}}
\newcommand{\p}{\varphi}
\newcommand{\cL}{{\cal L}}
\newcommand{\dphi}{\delta\varphi}
\newcommand{\dbj}{\delta{\bf j}}

\newcommand{\rred}[1]{{#1}}
\newcommand{\skp}[1]{{#1}}
\newcommand{\bblue}{}

\title{ Electron-Electron scattering and resistivity of ballistic multimode channels}

\author{ K. E. Nagaev and N. Yu. Sergeeva}
\affiliation{Kotelnikov Institute of Radioengineering and Electronics,  Mokhovaya 11-7, 
 Moscow, 125009 Russia}

\date{\today}

\begin{abstract}
We show that electron--electron scattering gives a positive contribution to the resistivity
of ballistic multimode wires whose width is much smaller than their length. This contribution is
not exponentially small at low temperatures and therefore may be experimentally observable. It scales
with temperature as $T^2$ for three-dimensional channels and as $T^{5/2}$ for 
two-dimensional ones.
\end{abstract}

\maketitle

\section{introduction}

Electron--electron scattering does not contribute to the resistivity of a 
homogeneous conductor with a pa\-ra\-bo\-lic spectrum. The reason is that the current density is proportional 
to the total momentum of electron gas, which is conserved in electron--electron collisions. If umklapp processes \cite{umklapp} are neglected, this 
contribution may be nonzero only in a presence of microscopic or macroscopic inhomogeneities. In 
particular, quantum-mechanical interference between electron-electron and electron--impurity scattering
results in a temperature-dependent correction to the conductivity.\cite{Altshuler,Zala} The electron--electron scattering is also known to affect the resistance of narrow channels with boundary scattering because it deflects electrons moving along the channel axis to the boundaries, where they can dissipate their momentum.\cite{Molenkamp} In short and wide ballistic contacts, the scattering in the electrodes also influences the resistance because the collisions change trajectories of electrons near an aperture in a diaphragm .\cite{MacDonald82,Nagaev08,Nagaev10} The electron-electron scattering is also known to influence the conductance of quantum contacts with saddle-point potential sharp enough to violate momentum conservation.\cite{Lunde09}

In all the above cases, the inhomogeneities are introduced into the system explicitly. However in finite-length conducting channels, the inhomogeneity is due to a
mere presence of electron reservoirs at the ends. In \cite{Lunde06}, Lunde et al. found that the contribution of
electron-electron scattering to the conductance of a two-channel quantum wire attached to reservoirs becomes nonzero at some specific relations of the Fermi wave vectors of electrons in different channels. More recently, a number of authors \cite{Lunde07,Micklitz10,Levchenko10,Levchenko11} addressed the effect of electronic collisions on the conductance of a single-channel quantum wire 
of a finite  length. As two-electron collisions do not affect the current in a strictly one-dimensional 
system because of momentum and energy conservation, this contribution is determined by triple electronic collisions, which involve an empty electronic state 
near the bottom of the band. Therefore it is exponentially small at low temperatures and moreover, it is zero for some scattering potentials, e.g. for the point-like one.  Importantly, this effect 
does not require an explicit presence of an inhomogeneity that can absorb electronic momentum. Instead,
this inhomogeneity is implicitly introduced through the boundary conditions at the ends of the wire, which
state that electrons coming to the wire from a reservoir have the same equilibrium distribution as in the reservoir. 

In this paper, we calculate the resistivity of a finite-length narrow multichannel ballistic conductor that  results from weak electron-electron scattering. Because the motion of electrons is also possible in the transverse direction, two-particle collisions affect the current despite the conservation laws. These 
collisions involve only electrons near the Fermi level, and therefore the effect is not exponentially small at low temperatures. 

The paper is organized as follows. In Sec. II we present the model and basic equations, in Sec. III we discuss the results, and Sec. IV presents the summary. Appendices A and B contain details of calculations for the 3D and 2D cases.

\begin{figure}[t]
 \includegraphics[width=8.5cm]{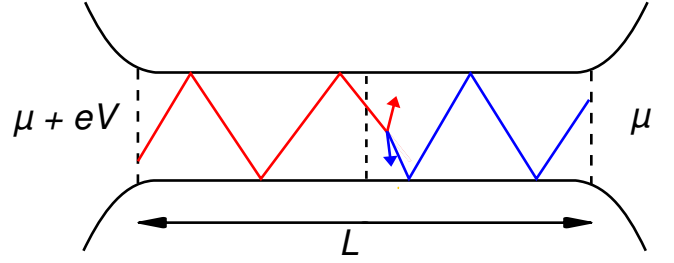}
\caption{\label{fig1} A sketch of the ballistic multimode wire with two electrons colliding half-way.
 The collision of the left-moving  electron (blue) with the right-moving one (red) prevents the left-mover from crossing the section of the wire shown by the dashed line in the middle and therefore affects the current flowing through it despite the momentum conservation. }
\end{figure}
\section{Model and basic equations}

Consider a  metallic wire of uniform cross-section connecting two massive electronic reservoirs. The length $L$ of the 
wire is much larger than its transverse dimension $a$ but much smaller than both the elastic and inelastic mean free paths. In its 
turn, $a$ is much larger than the Fermi wavelength $\lambda_F$. It is assumed that electrons are
specularly reflected at the boundaries of the wire, so they cannot transfer their longitudinal momentum to the lattice. In a two-dimensional (2D) case, such a wire could be realized by twisting up a 2D conducting strip to form a tube so that the boundary scattering could be eliminated at all.
We neglect here the effects of electron-electron scattering near the contacts of the wire with the reservoirs \cite{Nagaev08,Nagaev10}. The kinetic equation for the distribution function of electrons inside the wire is of the form \cite{Lunde06}
\be
 \bv\,\frac{\partial f}{\partial\br}
 =
 \hat{I}_{ee},
 \label{Boltz}
\ee
where the collision integral is given by
\begin{multline}
 \hat{I}_{ee}(\bp)
 =
 \alpha_{ee}\,\nu_d^{-2}
 \int\frac{d^dk}{(2\pi)^d}
 \int\frac{d^dp'}{(2\pi)^d}
 \int d^dk'\,
\\ {}\times
 \delta(\bp + \bk - \bp' - \bk')\,
 \delta( \eps_{\bp} + \eps_{\bk} - \eps_{\bp'} - \eps_{\bk'} )
\\ {}\times
 \Bigl\{
  [1 - f(\bp)]\,[1 - f(\bk)]\, f(\bp')\, f(\bk')\,
\\{}  -
  f(\bp)\, f(\bk)\, [1 - f(\bp')]\, [1 - f(\bk')]
 \Bigr\},
 \label{I_ee}
\end{multline}
$\alpha_{ee}$ is the dimensionless interaction parameter, $d=2$ or 3 is the dimensionality
of the system; $\nu_3=mp_F/\pi^2$ and $\nu_2=m/\pi$
are the three- and two-dimensional two-spin electronic densities of states ($\hbar=1$). The term with electric field
is omitted because we address the linear response and the voltage drop is included in the boundary 
conditions
\begin{align}
  f(p_x>0,x=0) &  = f_R(\eps_{\bp} - E_F), 
  \\
  f(p_x<0,x=L) &  = f_L(\eps_{\bp} - E_F), 
 \label{bound_cond}
\end{align}
where $x$ is the longitudinal coordinate, $f_R(\eps)=f_0(\eps-eV)$ and $f_L(\eps)=f_0(\eps) = 1/[1 + \exp(\eps/T)]$. The current through an 
arbitrary section of the conductor is given by an integral over the transverse coordinates
\be
 I=
 2e\int d^{d-1}r_{\perp}\,
 \int\frac{d^d p}{(2\pi)^d}\,
 v_x\, f(\bp,x,\br_{\perp}).
 \label{current}
\ee

Equation (\ref{Boltz}) may be solved iteratively in the collision integral. In the zero approximation, 
the electrons move inside the wire along broken lines while retaining their energy and longitudinal momentum
(see Fig. 1), so their distribution
depends solely on the reservoir from which the trajectory of an electron with a given momentum originates.
Therefore  the coordinate-independent  distribution is of the form 
\begin{multline}
 f^{(0)}(\bp,\br) = \Theta(p_x)\,f_R(\eps_{\bp}-E_F) 
\\{}
 + \Theta(-p_x)\,f_L(\eps_{\bp}-E_F)
 \label{f}
\end{multline}
and the zero-approximation two-spin conductance is just given  by the Sharvin formula 
\cite{Sharvin,Kulik77}
\be
 G_{03} = \frac{e^2 a^2 p_F^2}{4\pi^2},
 \qquad
 G_{02} = \frac{e^2 p_F a}{\pi^2}.
 \label{Sharvin}
\ee

To the first approximation in $\alpha_{ee}$, the correction to the distribution function 
$\delta f(\bp,\br)$ at a given point $\br$ of the conductor is readily obtained by integrating $\hat{I}_{ee}\{f^{(0)}\}$ along the trajectory of an electron with momentum $\bp$ that comes 
to point $\br$ from the corresponding reservoir. As $\hat{I}_{ee}\{f^{(0)}\}$ is coordinate-independent,
this integration is reduced to multiplying it by the time of motion inside the conductor
\begin{align}
 \delta f(p_x>0,\bp_{\perp},x)
 = \frac{x}{v_x}\,\hat{I}_{ee}(p_x>0,|p_{\perp}|),
\label{df_R}\\
   \delta f(p_x<0,\bp_{\perp},x)
 = \frac{L-x}{|v_x|}\,\hat{I}_{ee}(p_x<0,|p_{\perp}|).
\label{df_L}
\end{align}
Upon a substitution of Eqs. (\ref{df_R}) and (\ref{df_L}) into Eq. (\ref{current}) $v_x$
nicely drops off and taking into account the particle-number conservation by the collision
integral, one obtains an expression similar to that of \cite{Lunde07}
\be
 \delta I = 2eLa^{d-1} \int\frac{d^d p}{(2\pi)^d}\,
 \Theta(p_x)\,\hat{I}_{ee}(\bp),
 \label{dI2}
\ee
which tells us that $\delta I$ is proportional to the rate of change in the number of electrons
with $p_x>0$. 
One might think that it contradicts the momentum-conservation law, but this is 
not the case. Though collisions between electrons do not change their total momentum, they change their trajectories and hence may prevent some of them from passing through the cross-section of the conductor
at which the current is calculated (see Fig. 1).

Equation (\ref{df_R}) must be substituted into (\ref{dI2}). 
Because of a piecewise distribution (\ref{f}), it is convenient to present the result as a sum of
integrals over left-moving and right-moving electronic states.
Upon a decomposition of momentum integration into the integrations over the angles and
 over the energies, one obtains 
\begin{multline}
 \delta I = 2\frac{e a^{d-1} L \alpha_{ee}}{\nu_d^2}
 \sum_{\alpha\beta\gamma = (L,R)}
 \int d\eps \int d\eps' \int d\eps_1 \int d\eps_2\,
\\{}\times
 \delta(\eps + \eps' - \eps_1 - \eps_2)
\\{}\times
 F_{R\alpha\beta\gamma}(\eps,\eps'|\eps_1,\eps_2)\,
 B_{R\alpha\beta\gamma}(\eps,\eps'|\eps_1,\eps_2),
 \label{dI4}
\end{multline} 
where the indices $\alpha$, $\beta$, and $\gamma$ correspond
to left-moving (L) and right-moving (R) states.
The distribution-dependent factor is given by
\begin{multline}
 F_{R\alpha\beta\gamma}(\eps, \eps'| \eps_1, \eps_2)
\\
 =
  [1 - f_R(\eps)]\,[1 - f_{\alpha}(\eps')]\, 
  f_{\beta}(\eps_1)\, f_{\gamma}(\eps_2)\,
\\{}
  -
  f_R(\eps)\,f_{\alpha}(\eps')\, 
  [1 - f_{\beta}(\eps_1)]\,[1 - f_{\gamma}(\eps_2)]
\label{F}
\end{multline}
with $f_R(\eps) = f_0(\eps-eV)$ and $f_L(\eps) = f_0(\eps)$. The quantity
\begin{multline}
 B_{R\alpha\beta\gamma}(\eps,\eps'|\eps_1,\eps_2)
 =
 \int\frac{d^dp}{(2\pi)^d}
 \int\frac{d^dk}{(2\pi)^d}
 \int\frac{d^dp'}{(2\pi)^d}
\\{}\times
 \int d^2k'\,
 \delta(\eps_{\bp} - E_F - \eps)\,  \delta(\eps_{\bk} - E_F - \eps')\,
\\{}\times
 \delta(\eps_{\bp}'- E_F - \eps_1)\,\delta(\eps_{\bk}'- E_F - \eps_2)
 \Theta(p_x)\,\Theta(\sigma_{\alpha}\,k_x)\,
\\{}\times
 \Theta(\sigma_{\beta}\,p'_x)\,\Theta(\sigma_{\gamma}\,k'_x)\,
 \delta(\bp + \bk - \bp' - \bk')
 \label{B}
\end{multline}
with $\sigma_R=1$ and $\sigma_L=-1$ is the factor representing the phase space available for 
scattering. The quantity $B_{RRLL}$ is zero because of momentum conservation, and the
distribution-dependent factors $F_{RRRR}$, $F_{RLRL}$, and $F_{RLLR}$ vanish because of energy 
conservation. Hence only four terms in the sum (\ref{dI4}) containing 
odd numbers of left-moving and right-moving states are nonzero.
Note that $B_{\alpha\beta\gamma\delta}$ is symmetric with respect to permutations of
the first and second pairs of arguments and to permutations inside these pairs and 
also symmetric with respect to a simulataneous reversal of all 
its indices. Using these symmetry properties and linearizing $F_{R\alpha\beta\gamma}$ with respect
to the voltage, one may reduce the sum (\ref{dI4}) to a single term
\begin{multline}
 \delta I = -4\frac{e a^{d-1} L \alpha_{ee}}{\nu_d^2}\,\frac{eV}{T}
 \int d\eps \int d\eps' \int d\eps_1 \int d\eps_2\,
\\ {}\times
 \delta(\eps + \eps' - \eps_1 - \eps_2)\,
 \tilde{F}(\eps,\eps'|\eps_1,\eps_2)
 B(\eps,\eps'|\eps_1,\eps_2),
 \label{dI6}
\end{multline}
where
\begin{multline}
 \tilde{F}(\eps,\eps'|\eps_1,\eps_2)
 =[1 - f_0(\eps)]\,[1 - f_0(\eps')]\,
\\{}\times
 f_0(\eps_1)\,f_0(\eps_2).
 \label{F1}
\end{multline}

The quantity $B \equiv B_{RRRL}$ exhibits different behaviors for $d=3$ and $d=2$. In the three-dimensional case,
 it is dominated by collisions of electrons with total momentum $\bf Q$ of the order of $p_F$ and 
remains finite even if all the initial and final states lie exactly at the Fermi surface. This may be understood as follows. Consider two electrons
with momenta $\bp$ and $\bk$ lying exactly on the Fermi sphere that scatter into $\bp'$ and $\bk'$, respectively. Because of the conservation laws, the two initial and the two final states form a rectangle inscribed in the sphere and lying in a plane perpendicular to $\bf Q$, which is located at a distance $Q/2$ from its center (see Fig. 2). Depending on the direction and length of $\bf Q$, the plane $p_x=0$ may cut the rectangle into two parts and isolate two, one, or none of the vertices from it. The quantity $B_{RRRL}$ is determined by
momentum configurations with just one isolated vertex, which exist even in the zero-energy limit. To the lowest order in $\eps$, calculations give (see Appendix A) $B=C_3\, m^4 p_F$ where $C_3=0.222$, and the relative correction to the conductance is
\begin{figure}[t]
 \includegraphics[width=7cm]{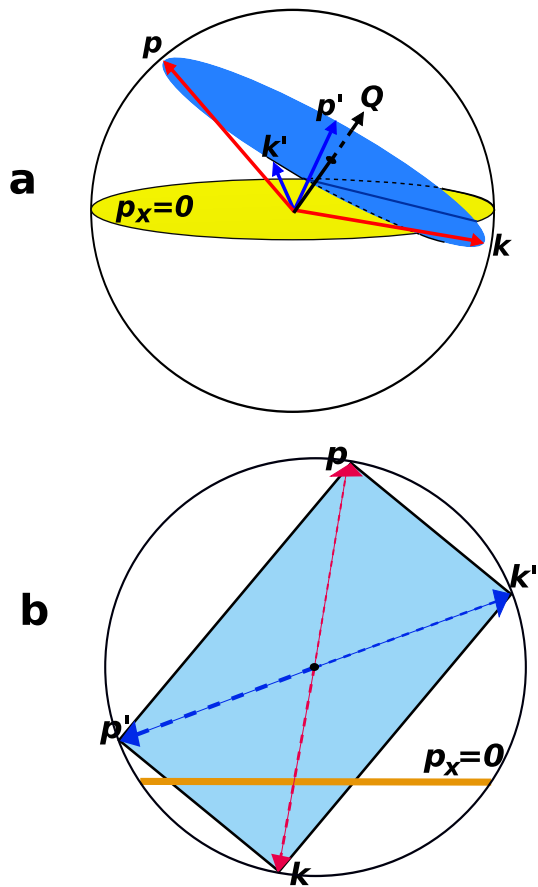}
 \caption{\label{fig2} A 3D Fermi surface and the four states involved in a collision
of two electrons. States $\bp$, $\bp'$ and $\bk'$ correspond to electrons moving from left to 
right whereas state $\bk$ corresponds to an opositely moving electron. (a) A 3D view of the Fermi sphere
 and (b) its section with the plane containing the four states with the trace of the plane $p_x=0$ (yellow line). The collision increases the number of right-movers by one.}
\end{figure}

\be
 \frac{\delta G_3}{G_{03}} = 
 -\frac{C_3}{3}\,
 \frac{L}{v_F}\,\frac{\alpha_{ee} T^2}{E_F}
 \sim -\frac{L}{l_{ee}},
 \label{dG3}
\ee
where $l_{ee}$ is the standard length of electron-electron scattering. This means that a significant
proportion of electron--electron collision contributes to the current.

In the two-dimensional case, $B$ vanishes at $\eps=\eps'=\eps_1=\eps_2=0$, and therefore the correction to the current is determined by collisions with small total momenta. Explicit calculations give (see Appendix B)
\be
 B(\eps,\eps'|\eps_1,\eps_2)=
 \frac{\sqrt{2}m^{3/2} T^{1/2}}{16\pi^6v_F^3}\,
 S\!\left( \frac{\eps - \eps'}{2T}, \frac{\eps_1 - \eps_2}{2T} \right)\!,
 \label{B4}
\ee
where
\be
 S(x,y) =
 \begin{cases}
  \sqrt{|x| + y} + \sqrt{y - |x|}, &  y>|x|   \\
  \sqrt{|x| + y},                  &  |y|<|x| \\
  0,                               &  y<-|x|.
 \end{cases}
 \label{S}
\ee 
The difference from the
three-dimensional case is readily seen in Fig. 3. All the four momenta may lie exactly 
at the Fermi circle only at ${\bf Q}=0$. In that case, the center of the rectangle 
coincides with the center of the Fermi circle and the line $p_x=0$ cuts it into two equal parts so 
that two vertices lie in the half-plane $p_x>0$ and the other two, in the half-plane $p_x<0$. If the
four momenta are located within $\delta\eps \sim T$ from the Fermi surface, a rectangle with one 
isolated vertex may be placed in this ring-shaped region only if all its vertices are located near 
the line $p_x=0$, where the portions of the ring at the opposite sides of the Fermi surface are almost
parallel. Therefore characteristic values of $Q$ that dominate $B$ are determined by the length at
which the bending of these portions is of the order of their width $\delta\eps/v_F$, i.e. $Q \sim \sqrt{p_F\eps/v_F}$. A substitution of Eqs. (\ref{B4}) and (\ref{S}) into (\ref{dI6}) gives
\be
 \frac{\delta G_2}{G_{02}}=
 -\frac{C_2}{2\pi^2}\,\frac{L}{v_F}\,
 \frac{\alpha_{ee} T^{5/2}}{E_F^{3/2}},
 \label{dG2}
\ee
where $C_2=2.23$.
This suggests that at low temperatures, $\delta G_2/G_{02}$ is smaller than $L/l_{ee}$ by a factor
$(T/E_F)^{1/2}$ and only a small fraction of electron--electron collisions affects the current.

The proportionality of $\delta G_3$ and $\delta G_2$ to $L$ is the result of the linear approximation
in $\alpha_{ee}$. In fairly long wires, this approximation breaks down. It is reasonable to believe that 
at $L\to\infty$ the corrections to the conductance saturate to a finite value as is in the case of a single-mode channel \cite{Micklitz10} and therefore present a boundary effect. 

\begin{figure}
 \includegraphics[height=7cm]{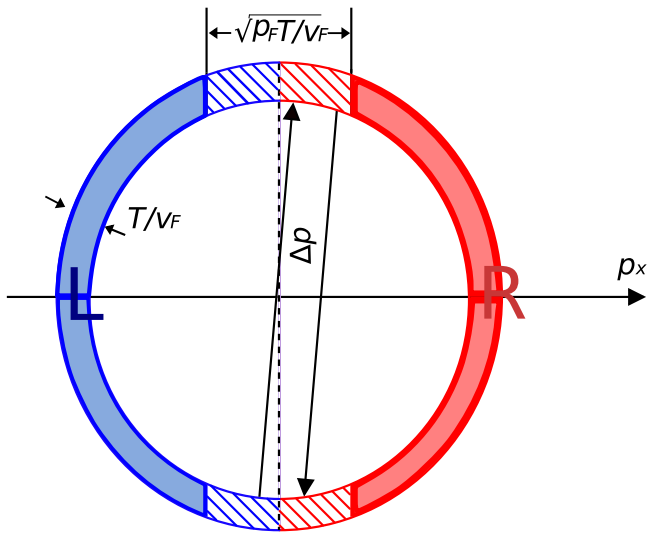}
 \caption{\label{fig3} A 2D Fermi surface and a scattering process that changes the number
 of right-moving electrons. The dashed areas show the parts of the surface that contribute to such 
processes.}
\end{figure}

\section{Discussion}

In contrast to the positive contribution to the conductance from the scattering in short and wide 
ballistic contacts \cite{MacDonald82,Nagaev08,Nagaev10}, the negative contribution in long and narrow wires does
not require an explicit transfer of electron momentum to a macroscopic obstacle. Instead, the excess
momentum is absorbed in the reservoirs just like the energy of injected electrons. The principal difference from the case of short contacts
is that  the collisions in a long and narrow channel involve only two type of electrons, 
left-movers and right-movers,which both contribute to the current with opposite signs. In contrast to this, in  wide and short contacts the collisions involve also a third type of electrons, by-passers. They by far outnumber the first two types, do not contribute to the current
in the absence of scattering but can be converted into a left-mover or right-mover upon a collision.  

The scattering 
mechanism in wires is more efficient in 3D than in 2D because the fraction of collisions contributing to the current increases with the dimensionality. Unlike this, the scattering contribution in wide contacts is smaller in 3D than in 2D because the concentration of injected nonequilibrium electrons falls off faster in 3D reservoirs.

The scattering contribution to the resistance may be well observable in GaAs heterostructures under realistic experimental condition. For an electronic density $n_s = 10^{11}$ cm$^{-2}$ and $T=4$ K 
high-mobility samples may have the elastic mean free path as large as 20 $\mu m$. For a conducting 
channel of that length and $\alpha_{ee} \sim 1$, the correction to the conductance should be about 10\%.
This correction can be isolated from the noninteracting Sharvin resistance (\ref{Sharvin}) by changing the length of the channel $L$ in a 2D electron gas
by means of electrostatic gates and isolating the component of the resistance proportional to $L$.

\section{Summary}

In summary, we have shown that two-electron collisions contribute to the electric resistivity of
multimode finite-length metallic wires even in the absence of umklapp processes. The finite contribution
to the resistivity is due to the breaking of translational symmetry of the system by the presence of reservoirs. This effect has a power-law temperature dependence and is more pronounced in 3D wires  than in
2D ones.

\begin{acknowledgments}
This work was supported by Russian Foundation for Basic Research, grants 10-02-00814-a and 11-02-12094-ofi-m-2011, by the program of Russian Academy of Sciences, the Dynasty Foudation, and by the Ministry of Education and Science of Russian Federation, contract  No 16.513.11.3066.
\end{acknowledgments}

\appendix
\section{Calculation of $B$ for the 3D case}

The phase space available for scattering (\ref{B}) may be rewritten in a form
\begin{multline}
 B(\eps,\eps'|\eps_1,\eps_2)=
 \int\frac{d^dQ}{(2\pi)^d}\,A_{RR}(\eps,\eps',{\bf Q})
\\{}\times
 A_{RL}(\eps_1,\eps_2,{\bf Q}),
 \label{B2}
\end{multline}
where ${\bf Q} = \bp + \bk = \bp' + \bk'$ is the total momentum of the colliding electrons and 
\begin{multline}
 A_{\alpha\beta}(\eps,\eps',{\bf Q})=
 \frac{1}{(2\pi)^d}\int d^dp \int d^dk\,
 \Theta(\sigma_{\alpha}\,p_x)\,\Theta(\sigma_{\beta}\,k_x)\,
\\{}\times
 \delta(\eps_{\bp}-E_F -\eps)\,\delta(\eps_{\bk} - E_F - \eps')\,
 \delta(\bp + \bk - {\bf Q}).
 \label{A}
\end{multline}
In 3D, explicit calculations give 
\begin{multline}
 A_{RR}(\eps,\eps', {\bf Q}) 
\\
 =
 \frac{2m^2}{(2\pi)^3Q}\,\Theta(\Delta)
 \int\limits_0^{\pi} d\p\,
 \Theta\!\left(
  \frac{Q_x}{Q_{\perp}}\,\frac{p^2_{\eps} - p^2_{\eps'} + Q^2}{\sqrt{\Delta(Q,\eps,\eps')}} - \cos\p
 \right)
\\{}\times
 \Theta\!\left(
  \cos\p - \frac{Q_x}{Q_{\perp}}\,\frac{p^2_{\eps} - p^2_{\eps'} - Q^2}{\sqrt{\Delta(Q,\eps,\eps')}}
 \right)
 \label{R-3D-1}
\end{multline}
and
\begin{multline}
  A_{RL}(\eps,\eps', {\bf Q}) 
 =
 \frac{2m^2}{(2\pi)^3Q}\,\Theta(\Delta)
\\{}\times
 \int\limits_0^{\pi} d\p\,
 \Theta\!\left(
  \frac{Q_x}{Q_{\perp}}\,\frac{p^2_{\eps} - p^2_{\eps'} - Q^2}{\sqrt{\Delta(Q,\eps,\eps')}} - \cos\p
 \right),
 \label{L-3D-1}
\end{multline}
where $p_{\eps}=\sqrt{2m(\eps+E_F)}$, $p_{\eps'}=\sqrt{2m(\eps'+E_F)}$, and
\begin{multline}
 \Delta(Q,\eps,\eps')
 =\Bigl[Q^2 - (p_{\eps} - p_{\eps'})^2\Bigr]\,
\\{}\times\Bigl[(p_{\eps}+p_{\eps'})^2 - Q^2\Bigr].
 \label{Delta}
\end{multline}
The calculations are greatly simplified if $\eps - \eps' \to 0$. By going to spherical coordinates
in (\ref{B2}) and introducing new integration variables 
$u = Q/2p_F$ and $v = Q^2\,\Delta^{-1/2}\cot\theta$ instead of the radial coordinate $Q$ and the polar angle
$\theta$, one arrives at an expression $B= C_3\, m^4 p_F$, where
\begin{multline}
 C_3= \frac{1}{2^4\pi^8}
 \int\limits_0^1 du\,u^2\,\sqrt{1 - u^2}
\\{}\times
 \int\limits_0^1 dv\,
 \frac{\arcsin v\, \arccos v}{(u^2 + v^2 - u^2 v^2)^{3/2}}
 =0.222.
\label{B3}
\end{multline}
\section{Calculation of $B$ for the 2D case}

Equations (\ref{B2}) and (\ref{A}) are also valid in the 2D case, but the explicit expressions
for $A_{RR}$ and $A_{RL}$ are now of the form
\begin{multline}
 A_{RR}(\eps,\eps',{\bf Q}) = \frac{m^2}{2\pi^2}\,\frac{\Theta(\Delta)}{\sqrt{\Delta(Q,\eps,\eps')}}
\\{}\times
 \Bigl\{
  \Theta\!\Bigl[
    (Q^2 + p_{\eps}^2 - p_{\eps'}^2)\,Q_x + |Q_y|\,\sqrt{{\Delta(Q,\eps,\eps')}}
  \Bigr]
\\{}\times
  \Theta\!\Bigl[
    (Q^2 - p_{\eps}^2 + p_{\eps'}^2)\,Q_x - |Q_y|\,\sqrt{{\Delta(Q,\eps,\eps')}}
  \Bigr]
\\{}+
  \Theta\!\Bigl[
    (Q^2 + p_{\eps}^2 - p_{\eps'}^2)\,Q_x - |Q_y|\,\sqrt{{\Delta(Q,\eps,\eps')}}
  \Bigr]\,
\\{}\times
  \Theta\!\Bigl[
    (Q^2 - p_{\eps}^2 + p_{\eps'}^2)\,Q_x + |Q_y|\,\sqrt{{\Delta(Q,\eps,\eps')}}
  \Bigr]
 \Bigr\},
 \label{A_RR-2D}
\end{multline}
\begin{multline}
 A_{RL}(\eps_1,\eps_2,{\bf Q}) = \frac{m^2}{2\pi^2}\,\frac{\Theta(\Delta)}{\sqrt{\Delta(Q,\eps_1,\eps_2)}}
\\{}\times
 \Bigl\{
  \Theta\!\Bigl[
    (Q^2 + p_{\eps1}^2 - p_{\eps2}^2)\,Q_x + |Q_y|\,\sqrt{{\Delta(Q,\eps_1,\eps_2)}}
  \Bigr]
\\{}\times
  \Theta\!\Bigl[
    (p_{\eps1}^2 - p_{\eps2}^2 - Q^2)\,Q_x + |Q_y|\,\sqrt{{\Delta(Q,\eps_1,\eps_2)}}
  \Bigr]
\\{}+
  \Theta\!\Bigl[
    (Q^2 + p_{\eps1}^2 - p_{\eps2}^2)\,Q_x - |Q_y|\,\sqrt{{\Delta(Q,\eps_1,\eps_2)}}
  \Bigr]\,
\\{}\times
  \Theta\!\Bigl[
    (p_{\eps1}^2 - p_{\eps2}^2 - Q^2)\,Q_x 
\\{}- |Q_y|\,\sqrt{{\Delta(Q,\eps_1,\eps_2)}}
  \Bigr]
 \Bigl\}.
 \label{A_RL-2D}
\end{multline}
The integral (\ref{B2}) is conveniently calculated by going to radial coordinates. As Eqs.
(\ref{A_RR-2D}) and (\ref{A_RL-2D}) depend on the $|Q_y|/Q_x$ ratio only through $\Theta$-functions,
it takes up a form
\begin{multline}
 B(\eps,\eps'|\eps_1,\eps_2)
\\{} =
 \frac{m^4}{16\pi^6}
 \int\limits_0^{2p_F} dQ\,
 \frac{ Q\,\delta\p(Q) }
 {\sqrt{\Delta(Q,\eps,\eps')\,\Delta(Q,\eps_1,\eps_2)}},
 \label{B3-2D}
\end{multline}
where $\delta\p(Q)$ is the angular width of the domain of integration defined by the theta-functions in 
Eqs. (\ref{A_RR-2D}) and (\ref{A_RL-2D}). The integral is dominated by small values of $Q \sim \sqrt{p_F\eps/v_F} \ll p_F$, so that
\be
 \Delta(Q,\eps,\eps') \approx \Delta(Q,\eps_1,\eps_2)
 \approx 4p_F^2 Q^2.
\ee
The integration results in Eqs. (\ref{B4}) and (\ref{S}).

\end{document}